\documentclass[11pt,twoside]{article}
\usepackage{./asp2010}
\aspvolume{480}
\aspcpryear{2013}
\aspvoltitle{Structure and Dynamics of Disk Galaxies}
\aspvolauthor{Marc S.~Seigar and Patrick Treuhardt}
\resetcounters

\begin{document}

\title{The Sagittarius Impact on Light and Dark Structure in the Milky Way}
\author{Chris W. Purcell$^1$
\affil{$^1$Department of Physics and Astronomy, West Virginia University}
}

\begin{abstract}
It is increasingly apparent that common merger events play a large role in the evolution of disk galaxies at all cosmic times, from the wet accretion 
of gas-filled dwarf galaxies during the era of peak star formation, to the collisions between large, dynamically-advanced spiral galaxies and 
their dry companion satellites, a type of interaction that continues to influence disk structure into the present day.  We also live in a large spiral galaxy 
currently undergoing a series of impacts from an infalling, disrupting dwarf galaxy.  As next-generation astrometry proposes to place our understanding of 
the Milky Way spiral structure on a much firmer footing, we analyze high-resolution numerical models of this disk-satellite interaction in order to assess 
the dynamical response of our home Galaxy to the Sagittarius dwarf impact, and possible implications for experiments hoping to directly detect 
dark matter passing through the Earth.
\end{abstract}

Throughout this proceeding, we discuss the results initially presented in \citet{purcell_etal11} and \citet{purcell_etal12}, in which we perform 
high-resolution $N$-body experiments involving a dwarf galaxy model with parameters similar to those indicated by observational constraints 
on the Sagittarius (Sgr) progenitor \citep{niederste-ostholt_etal10}, as it is accreted into a Milky Way-like system that is initially a stellar disk in 
equilibrium with a host halo.  Following an orbital path and producing tidal debris consistent with the phase-space characteristics of the observed 
Sgr stream, the infalling subhalo impacts the disk twice in the recent Galactic past, with a third pericenter presently occurring as Sgr approaches 
the southern face of the Milky Way disk.  Each of these disk crossings results in the emergence of swing-amplified global spirality modes, such 
that the present-day Galaxy resembles a multi-armed spiral structure with intermediate-scale arms \citep[see the discussion of spirality categorization 
in][]{binney_tremaine08}.

\begin{figure}[t]
\begin{center}
\includegraphics[width=5.2in]{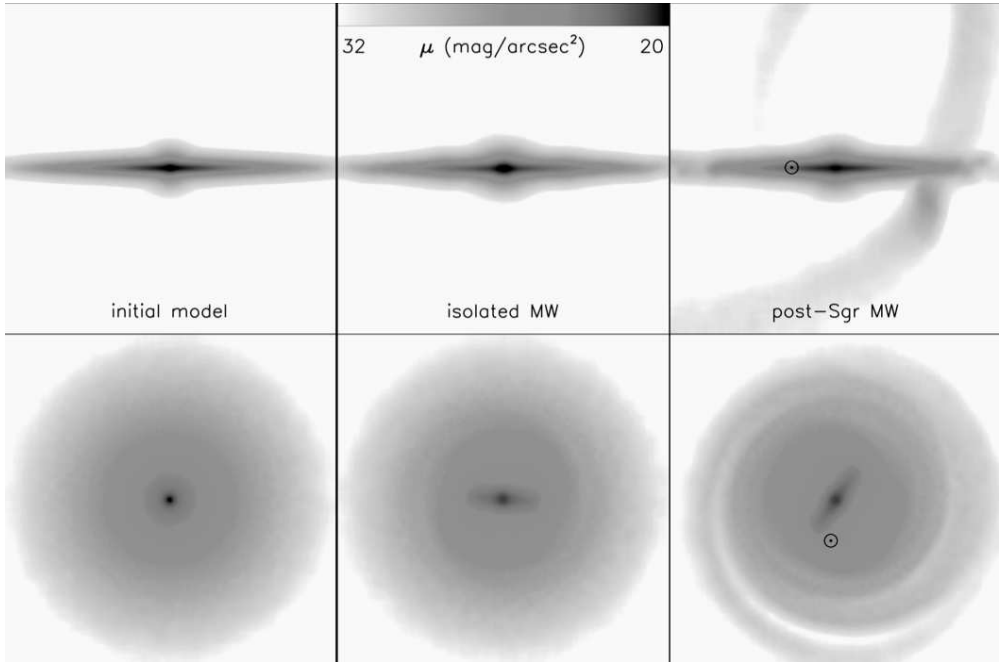} 
\end{center}
\caption{Surface brightness maps (assuming a stellar mass-to-light ratio M*/L = 3) for the initial state of our primary Milky Way model 
({\em left} panels) and the endstate following the infall of the {\em Light Sgr} dwarf galaxy ({\em right} panels). In the {\em center} panels, 
we show the result of secular evolution in the initial model for an equivalent timespan of 2.65 Gyr; note that the Sgr interaction causes the 
widespread emergence of discernible spiral arms and greatly influences the evolution of the central bar. In the right panels, the orientation 
of the satellite's orbital plane allows us to define the position of the Sun (marked by the solar symbol).}
\label{fig:disk}
\end{figure}

\begin{figure}[t]
\begin{center}
\includegraphics[width=5.2in]{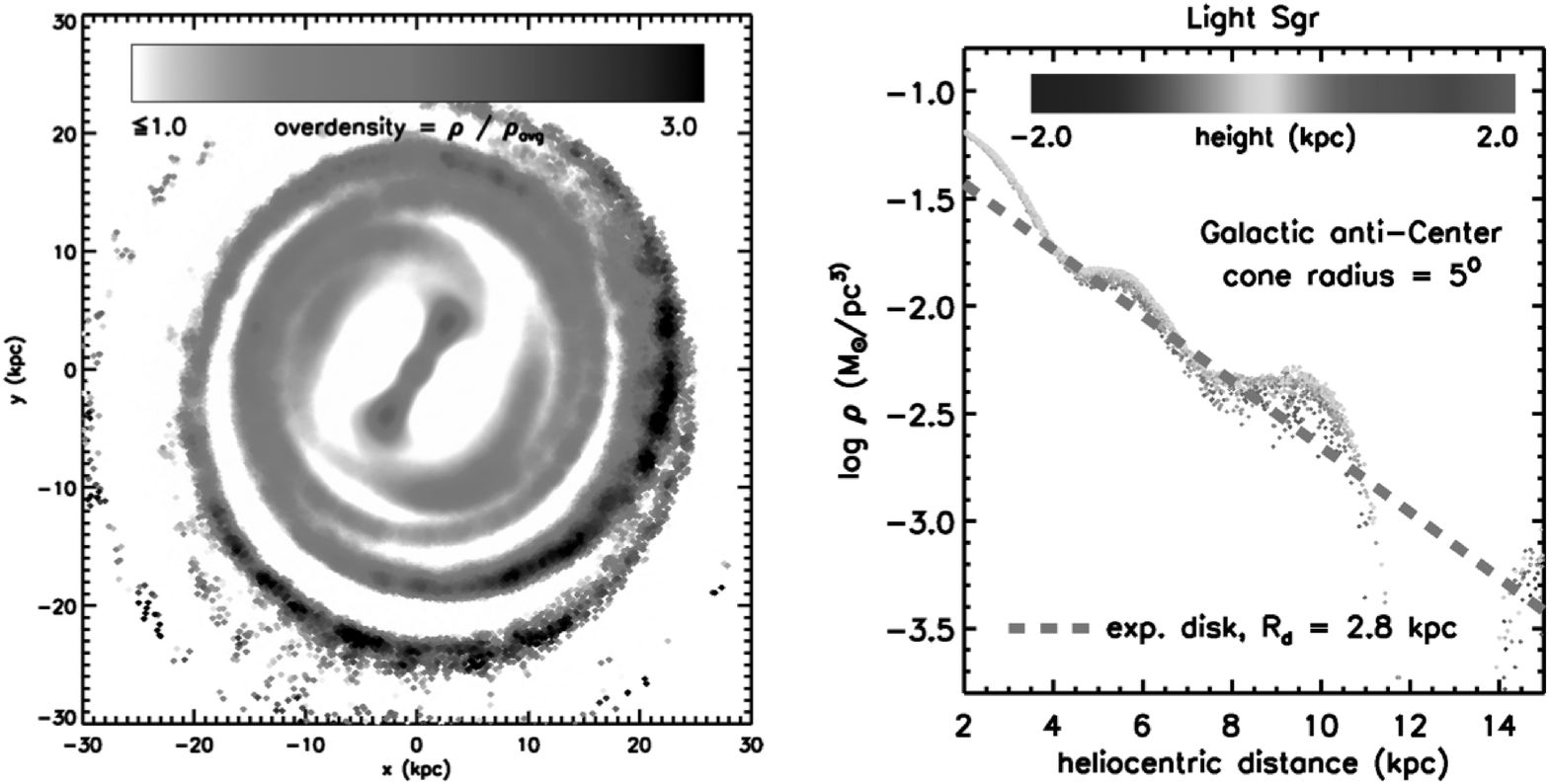} 
\end{center}
\caption{Visualization of the global disk structure in our {\em Light Sgr} simulation. The left panel presents an over-density map of the present-day 
stellar disk and the right panel shows the local density measured in a 5-degree cone emerging from the solar position and directed toward the Galactic anti-Center.}
\label{fig:spiral}
\end{figure}

The massive Sgr progenitor unfurls into a thick polar stream wrapping around the Milky Way, not only in stellar material but also in the attendant dark 
matter stripped from the satellite as it is accreted into the Galactic mass.  Being from a much dynamically hotter and larger distribution initially, the 
dark tidal debris cuts a much wider path across the Milky Way disk than the stellar Sgr stream, but is otherwise similarly distributed in phase space 
at the present day.  Serendipitously, the leading tidal arm of the disrupting Sgr system is falling down onto the northern face of the Galactic disk somewhere 
near the solar neighborhood; one robust conclusion of our work is that there must be Sagittarius dark matter moving coherently through the solar neighborhood 
at relatively high speeds, and that this feature in the dark matter speed distribution produces a significant phase shift in the expected annual modulation of 
nuclear-recoil events involving dark matter particles.

\section{Dynamical Response of the Milky Way to the Sagittarius Impact}

We visualize the stars associated with the primary Milky Way model in the left panels of Figure~\ref{fig:disk} and the same system evolved in 
isolation for 2.65 Gyr (the time of the longest Sgr orbit simulation) in the middle panels. We note that these initial conditions are quite stable 
during secular evolution, though a weak bar does begin to emerge after $\sim 2$ Gyr of evolution.  Each of the modeled Sgr progenitors experiences 
two disk crossings and approaches a third, corresponding to the three orbital pericenters (as indicated by shaded bands). For both models, 
the first impact occurs in the outer disk at $R_{GC} \sim 30$~kpc and is responsible for most of the dark matter mass loss in the satellite but very little 
stellar mass loss. The second occurs $\sim0.9$ Gyr later at $R_{GC} \sim 15$~kpc and is responsible for liberating the majority of the stellar stream debris. 
The final disk crossing is occurring at the present day, about $\sim1.8$ Gyr after the first disk crossing. We note that the impact times and radii for 
our models are quite similar to those presented in past work aimed at reproducing the Sgr stream in great detail, although these past models have not 
included a dark matter component to the massive Sgr galaxy \citep[see, {\em e.g.}][]{law_majewski10}.

The left panel of Figure~\ref{fig:spiral} visualizes the global disk response to the Sgr interaction in the {\em Light Sgr} model by using a face-on disk 
rendering color-coded by ratio of the local density to the radial/vertical average at each point. Under-densities are white in this depiction, while 
over-densities are colored.  The images reveal a prominent bar as well as a complicated array of spirality that reveals itself in the nearby anti-Center 
as overdensities rotating coherently with the Milky Way disk. The right panel depicts the local density as determined within a 
5-degree solid angle cone along the plane as a function of radius from the solar position, in the direction of the anti-Center region.

The disks in our simulations develop outer arcs of material generated in association with each disk crossing. These evolved outer wrappings are loosely 
wound and resemble rings. One of the predicted arcs, at about 10 kpc from the Sun, is reminiscent of the low-latitude Milky Way feature known as the 
Monoceros ring. Though the Monoceros ring is often considered to be the leftover tidal stream from a now-defunct dwarf satellite galaxy \citep{penarrubia_etal05}, 
some observational evidence has suggested that the Monoceros ring could be a feature of the Milky Way itself \citep{momany_etal06}. Previous theoretical 
work has suggested that a past encounter with some previously unidentified massive satellite could have produced the Monoceros ring as the outcome of a 
disk impact \citep{kazantzidis_etal08}, and that these structures are long-lived during the impact itself \citep{purcell_etal09}. We specifically identify the Sgr 
progenitor as the likely candidate for the impact that molded the Monoceros ring from the Milky Way disk, as the induced spiral arms detached from the outer 
Galactic plane and began to oscillate vertically over a range of $5-10$~kpc (as shown in the right panel of Figure~\ref{fig:spiral}).

\begin{figure}[t]
\begin{center}
\includegraphics[width=5.2in]{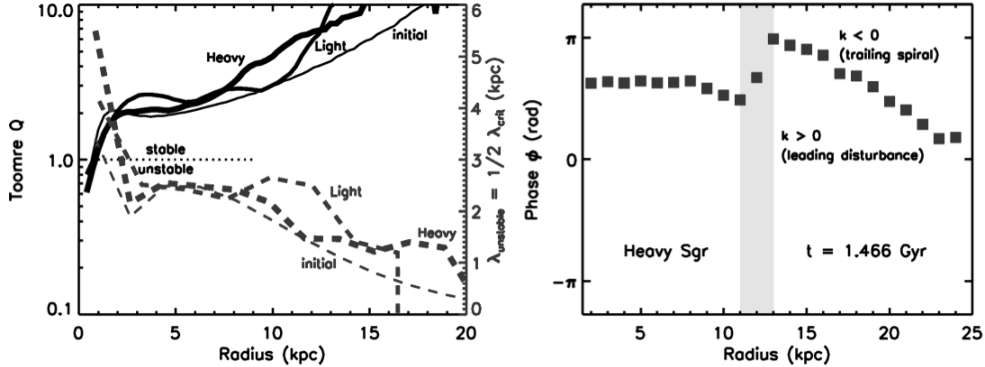} 
\end{center}
\caption{{\em Left} panel: the Toomre stability parameter Q and the critical wavelength associated with scales most susceptible to instability. 
Where necessary, we set the mode number $m=2$ for two-armed spiral and bar structure. Note that the initial Milky Way model, as well as 
both endstate disks from the {\em Light Sgr} and {\em Heavy Sgr} infall models, are globally stable with $Q > 1$ at all radii where disk particles 
dominate the stellar density. Small-scale perturbations with wavelengths less than roughly 2-3 kpc travel easily through all of these disks. 
{\em Right} panel: the phase (or shape function) $\phi$ as a function of radius, during the second pericentric approach/impact of the {\em Heavy Sgr} 
satellite, and annotated regionally with values of radial wavenumber $k = d\phi / dR$. Near the radius of disk impact (shaded area), a leading 
density disturbance emerges due to the gravitational perturbation of the interloping subhalo, and this feature is quickly sheared by differential 
rotation into the generally-trailing spiral structure of the post-infall disk as time evolves.}
\label{fig:modes}
\end{figure}

In order to quantify the presence of spiral arms in the resultant disk of the fiducial infall model, we have examined the Fourier amplitudes of the final 
density distribution, finding significant power in the $m=2$ and $m=4$ modes (corresponding to two-armed and four-armed morphologies respectively), 
with comparatively negligible power in the first two odd modes relative to axisymmetry. This is broadly consistent with analysis of satellite impacts onto 
the Milky Way \citep{chakrabarti_blitz09}, which finds that satellites with mass ratio $\sim1:50$ and orbital pericenter distances $\sim 20$ kpc do induce power 
in the first few Fourier modes, with comparatively greater amplitude in the $m=2$ mode. Analysis of the Toomre stability $Q$ parameter 
\citep[which similarly assesses axisymmetric disk instability, again see {\em e.g.}][]{binney_tremaine08} and associated critical wavelength $\lambda_{\mathrm{crit}}$ 
indicates that our fiducial Milky Way model is globally stable against long-wavelength perturbations, but is susceptible to short-wavelength modes on small 
scales, as demonstrated by Figure~\ref{fig:modes}.

\section{Dark Debris from Sagittarius on Earth}

There are particularly interesting consequences of the Sagittarius impact on experiments designed to test the local distribution of weakly-interacting 
dark matter particles via nuclear-recoil rates.  Previous numerical studies that have addressed the issue of direct detection and 
velocity distributions in high-resolution $N$-body simulations have largely been limited 
to analyses of simulations including only dark matter.  These simulations 
do not account for the Milky Way disk and its cosmological growth and evolution 
to the present day.  Such experiments represent only individual samples of the 
broad statistical ensemble of merger histories that lead to the formation of a 
Milky Way-sized dark matter halo.  Consequently, these simulated halos lack 
specific structures known to exist in the Milky Way, such as the ongoing 
Sagittarius accretion and tidal stream, or other features that may have yet to be discovered.  
Therefore, the problem of mapping the results of 
cosmological numerical simulations onto specific predictions for the actual 
Milky Way halo, especially at the solar neighborhood, is a distinct challenge.

\begin{figure}[t]
\begin{center}
\includegraphics[width=2.6in]{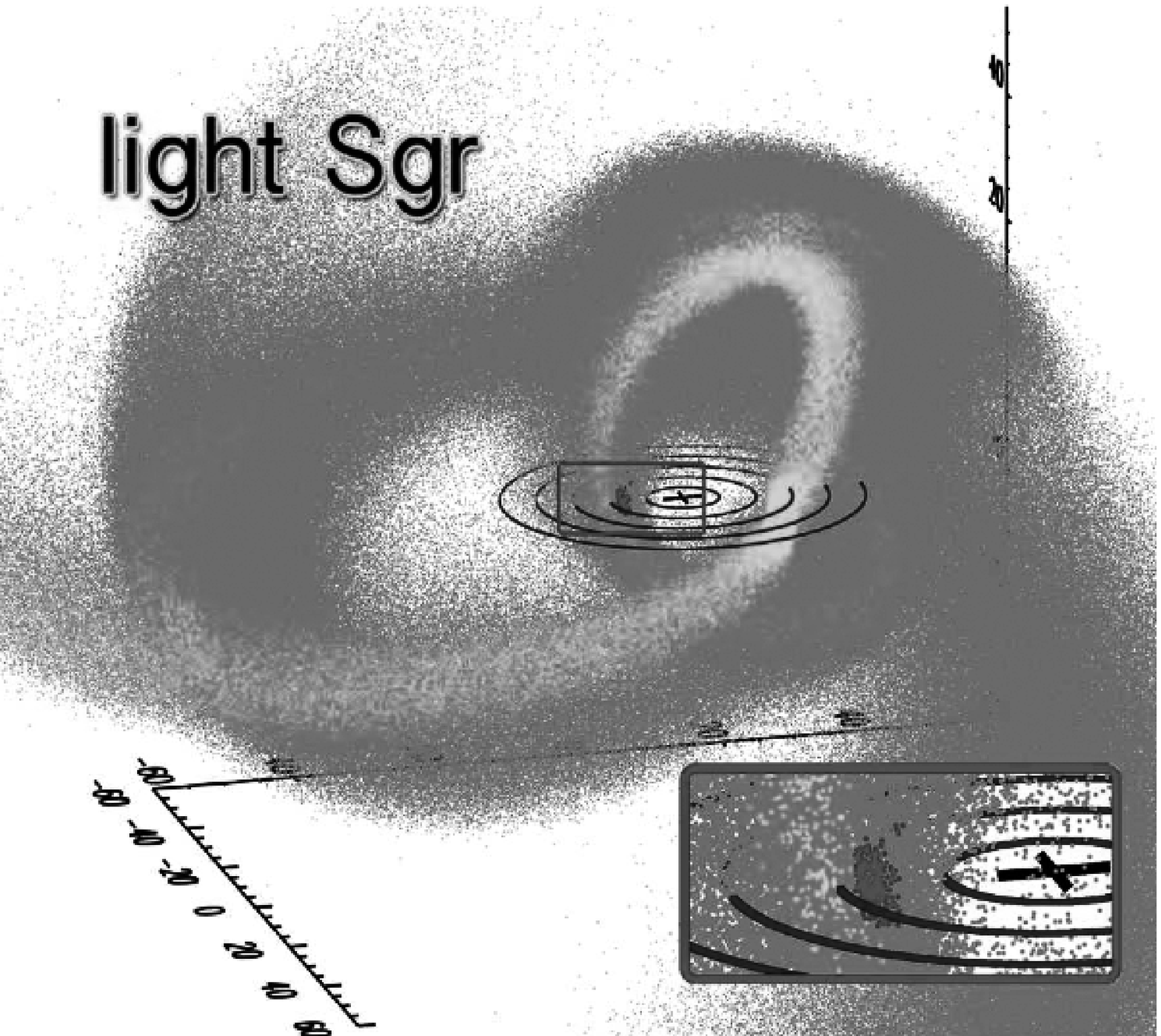} 
\includegraphics[width=2.6in]{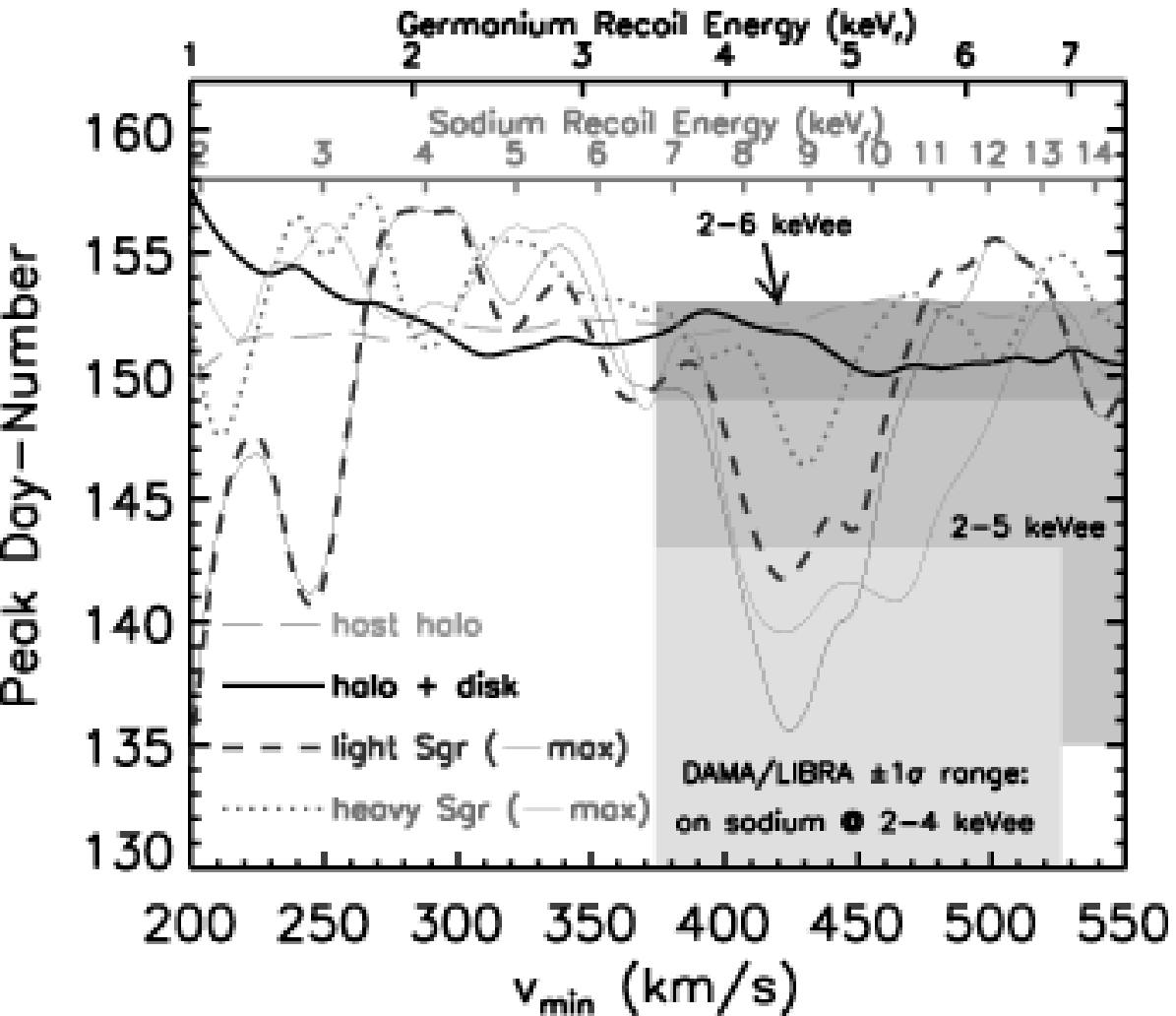} 
\end{center}
\caption{{\em Left} panel: distributions of stars and dark matter from the disrupting Sagittarius dwarf satellite galaxy (light and dark particles, respectively), 
in our {\em Light Sgr} model. In this perspective, the Milky Way disk plane is denoted by concentric blue rings at 5-kpc radial intervals, to a Galactocentric 
distance of 25 kpc.  The darkest particles appearing in the inset panel are in the solar neighborhood as defined in the text.
{\em Right} panel: Annual modulation phase as a function of recoil energy and the minimum necessary velocity $v_{\mathrm{min}}$; the shaded regions 
represent the $\pm1\sigma$ ranges in the modulation peaks determined by DAMA/LIBRA for scattering on sodium, 
{\em i.e.} peak day-numbers of 136, 142, 146 $\pm 7$ for recoil energy bins $E_r = 2-4,5,6$~keVee.  Where present, thin colored lines represent the maximum 
possible signal induced by each Sgr model.}
\label{fig:dark}
\end{figure}

Some observational measurements of the tidally-stripped stars suggest that the 
leading {\em stellar} arm of the Sgr tidal stream may fall several kiloparsecs away from the 
solar position in the Galactic plane (e.g., \citep{belokurov_etal06,seabroke_etal08}); although recent 
detections of coherently-moving stellar populations in the solar neighborhood seem to rule out large-scale 
flows with Sagittarius-like vertical velocities \citep{helmi_etal99,re_fiorentin_etal11}, percent-level streams outside one 
or two kiloparsecs from the Sun are poorly constrained and difficult to disentangle using tracer-sampling techniques.  
Observationally-viable models for the Sgr infall exhibit several interesting features 
relevant to contemporary and future direct search experiments, because the models we present here 
treat the Sgr dwarf in a cosmological context, in the sense that we assume 
that the Sagittarius galaxy is itself embedded in a dark matter halo as it merges with the 
Milky Way.  Utilizing simulations of the Sagittarius infall, we emphasize that although the primary {\em stellar} component of the Sagittarius debris stream may be 
several kiloparsecs away, the associated {\em dark matter} stream is always significantly 
more spatially extended than the stellar stream (as shown in Figure~\ref{fig:dark}) and 
more coherent in velocity space.  In fact, our models illustrate that the stellar and 
dark matter streams need not even be spatially concentric in realistic models.  In other words, 
the peak of the stellar density of the stream may be displaced from the peak in dark matter 
density associated with the stream by several kiloparsecs.  Both of these properties of 
simulated Sagittarius infall models suggest that it is not necessary for the solar neighborhood 
to be closer than several kiloparsecs away from the primary stellar stream of Sagittarius debris 
in order for the stream to affect significantly dark matter detection rates.  We use the dark matter 
streams realized in our simulations to model the dark matter velocity distribution in the solar 
neighborhood.  Indeed, we find that the near coincidence of the solar neighborhood and the debris 
of the Sagittarius satellite's dark matter halo creates a peak in the high-end of the dark matter 
velocity distribution.  These Sgr particles can lead to significant effects on the energy 
dependence of dark matter detection rates, as well as the amplitude and phase of the annual 
modulation of event rates, particularly for relatively low-mass dark matter candidates ($m_{\chi} \lesssim 20~\mathrm{GeV}$).

The geometry of the Sagittarius impact on the solar system 
causes the signal from the Sagittarius stream to peak during 
Northern winter, in agreement with previous studies 
\citep{freese_etal04,freese_etal05,savage_etal06,savage_etal07,savage_etal09} 
(as well as the general study of debris flows in 
\citet{vergados_2012}).  Our models of Sagittarius 
yield recoil energy-dependent shifts in peak day number of 
between $\sim 5$~and $25$ days earlier in the year than the SHM peak day-number of $152.5$.  
Both DAMA/LIBRA and CoGeNT have indicated a similar behavior, 
with both experiments finding trends between peak day-number and 
recoil energy \citep{bernabei_etal11,aalseth_etal11}; however, 
the observational error in the peak remains on the order of a few days 
and the uncertainty in simulation programs that aim to model Sagittarius 
remain significant.  Nevertheless, our models suggest such a shift 
is reasonable given contemporary knowledge of Sagittarius debris structure.  
Exploiting this signature, in particular, to help identify 
dark matter or use dark matter searches to perform WIMP 
astronomy will benefit greatly from future 
low-threshold detectors with improved energy resolution, 
such as those being considered by many projects \citep{texono,bruch2010,pyle_etal12}.  

The effects of Sagittarius that we describe in this manuscript may be relevant to 
dark matter searches generally.  However, if the dark matter is 
indeed relatively light ($m_{\chi} \lesssim 20$~GeV as we have assumed in 
our illustrative examples), the effects of Sagittarius debris on 
scattering rates are particularly important because of the large 
relative speed of the debris stream at the Earth.  In either case, 
future direct search experiments may probe such signatures, though a 
future generation of low-threshold detectors with fine energy 
resolution \citep{texono,bruch2010,pyle_etal12,graham_etal12}
may be necessary in the event that the dark matter mass falls in this 
lower range.  In either case, our analysis suggests that the effects of Sagittarius 
debris on direct search experiments {\em will not} be negligible given 
contemporary limits on the position of the Sagittarius {\em stellar} stream.  
In the far future, the features induced by Sagittarius debris may be among the early 
measurements to be made in an era of WIMP astronomy with large direct search rates \citep{peter2011}.

\acknowledgements CWP would like to thank James Bullock and Andrew Zentner for their continued 
involvement with the Sagittarius-impact projects, and the University of Arkansas, Little Rock for the 
organization of this conference as the impetus for interesting discussions with Robert Benjamin, 
Daniel Kennefick, Jerry Sellwood, among many others.

\bibliography{purcellc}

\end{document}